\documentclass[aps,prl,superscriptaddress,preprintnumbers,twocolumn,groupedaddress,showpacs]{revtex4}

\usepackage{graphicx}

\newcommand{\als}{\alpha_{\mathrm{s}}}
\newcommand{\MSbar}{\overline{\mathrm{MS}}}
\def\mathswitchr#1{\relax\ifmmode{\mathrm{#1}}\else$\mathrm{#1}$\fi}
\newcommand{\GF}{\mathswitch {G_\mu}}
\newcommand{\sw}{\mathswitch {s_\Pw}}

\newcommand{\rT}{{\mathrm{T}}}
\newcommand{\rj}{\mathrm{j}}
\newcommand{\gammainduced}{\mbox{\scriptsize $\gamma$-induced}}

\newcommand{\ppjjh}{\Pp\Pp\to\PH+2\mathrm{jets}+X}

\newcommand{\PH}{\mathswitchr H}
\newcommand{\Pw}{\mathswitchr w}
\newcommand{\PW}{\mathswitchr W}
\newcommand{\PZ}{\mathswitchr Z}
\newcommand{\Pp}{\mathswitchr p}

\def\mathswitch#1{\relax\ifmmode#1\else$#1$\fi}

\newcommand{\MH}{\mathswitch {M_\PH}}
\newcommand{\MW}{\mathswitch {M_\PW}}
\newcommand{\MZ}{\mathswitch {M_\PZ}}

\newcommand{\GeV}{\unskip\,\mathrm{GeV}}

\newcommand{\fb}{\unskip\,\mathrm{fb}}

\def\refeq#1{\mbox{(\ref{#1})}}

\def\reffi#1{\mbox{Figure~\ref{#1}}}

\def\refta#1{\mbox{Table~\ref{#1}}}

\def\citere#1{\mbox{Ref.~\cite{#1}}}
\def\citeres#1{\mbox{Refs.~\cite{#1}}}

\begin{document}

\preprint{MPP-2007-82, PSI-PR-07-03}
\title{\boldmath{
Strong and electroweak corrections to the 
production of Higgs+2jets via weak interactions at the LHC}}

\author{M.~Ciccolini}
\affiliation{Paul Scherrer Institut, W\"urenlingen und Villigen,
CH-5232 Villigen PSI, Switzerland}

\author{A.~Denner}
\affiliation{Paul Scherrer Institut, W\"urenlingen und Villigen,
CH-5232 Villigen PSI, Switzerland}

\author{S.~Dittmaier}
\affiliation{Max-Planck-Institut f\"ur Physik
(Werner-Heisenberg-Institut), D-80805 M\"unchen, Germany}

\date{\today}

\begin{abstract}
  Radiative corrections of strong and electroweak interactions are
  presented at next-to-leading order for the production of a Higgs
  boson plus two hard jets via weak interactions at the LHC.  The
  calculation includes all weak-boson fusion and quark--antiquark
  annihilation diagrams as well as the corresponding interferences.
  The electroweak corrections, which are discussed here for the first
  time, reduce the cross sections by $5\%$, and thus are of the same
  order of magnitude as the QCD corrections.  As argued in previous
  papers, where $s$-channel diagrams and interferences were neglected,
  the QCD corrections connected to interference effects are small.

\end{abstract}

\maketitle

\section{Introduction}

The production of a standard Higgs boson in association with two hard
jets is a cornerstone in the Higgs search both in the ATLAS
\cite{Asai:2004ws} and CMS \cite{Abdullin:2005yn} experiments at the
LHC for the Higgs mass range between 100 and $200\GeV$, which is
favoured by the global Standard Model fit to electroweak (EW)
precision data.

The production of Higgs+2jets receives two kinds of contributions at
hadron colliders.  The first type, where the Higgs boson couples to a
weak boson that links two quark lines, is dominated by squared $t$-
and $u$-channel-like diagrams and known as the ``vector-boson fusion''
(VBF) channel.  The hard jet pairs have a strong tendency to be
forward--backward directed in contrast to other jet production
mechanisms, offering a good background suppression
(transverse-momentum and rapidity cuts on jets, jet rapidity gap,
central-jet veto, etc.).  Upon applying appropriate event selection
criteria (see
e.g.~\citeres{Barger:1994zq,Rainwater:1997dg,Rainwater:1998kj,
  Rainwater:1999sd,DelDuca:2006hk} and
\citeres{Spira:1997dg,Djouadi:2005gi} for more references) it is
possible to sufficiently suppress background and to enhance the VBF
channel over the second $\PH$+2jets mechanism that mainly proceeds via
strong interactions.  In this second channel the Higgs boson is
radiated off a heavy-quark loop that couples to any parton of the
incoming hadrons via gluons \cite{DelDuca:2001fn,Campbell:2006xx}.
According to a recent estimate \cite{Acosta:2007it} hadronic
production contributes about $4{-}5\%$ to the Higgs+2jets events for a
Higgs mass of $120\GeV$ after applying VBF cuts.  A next-to-leading
order (NLO) analysis \cite{Campbell:2006xx} of this contribution shows
that its residual scale dependence is still of the order of 35\%.

Higgs production in the VBF channel is a pure EW process in leading
order (LO) involving only quark and antiquark parton distributions.
Approximating the cross section by $t$- and $u$-channel diagrams only
(without interference), because $s$-channel diagrams and interferences
are rather suppressed, the corresponding NLO QCD corrections reduce to
vertex corrections to the weak-boson--quark coupling. Explicit NLO QCD
calculations in this approximation
\cite{Han:1992hr,Spira:1997dg,Figy:2003nv,Figy:2004pt,Berger:2004pc}
confirm the expectation that these QCD corrections are quite small,
because they are shifted to the parton distribution functions (PDF)
via QCD factorization to a large extent. The resulting QCD corrections
are of the order of $5{-}10\%$ and reduce the remaining factorization
and renormalization scale dependence of the NLO cross section to a few
per cent.

In this paper we complete the previous NLO calculations for the VBF
channel in two respects. Firstly, we add the complete NLO EW
corrections, and secondly we include all interferences in the QCD
corrections.
While all interferences are negligibly small,
as expected, the EW corrections are of the same size as the 
QCD corrections and thus phenomenologically relevant.

\section{Details of the NLO calculation}

\begin{figure}
{\includegraphics[bb=125 665 330 715, width=.42\textwidth]{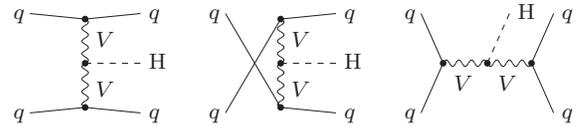}}
\vspace*{-1em}
\caption{Topologies for $t$-, $u$-, and $s$-channel contributions to
$qq\to qq\PH$ in LO, where $q$ denotes any quark or antiquark
and $V$ stands for W and Z~bosons.}
\label{fig:LOtops}
\end{figure}

At LO, the production of Higgs+2jets via weak bosons receives
contributions from the partonic processes $qq\to\PH qq$,
$q\bar{q}\to\PH q\bar{q}$, $\bar{q}\bar{q}\to\PH\bar{q}\bar{q}$. For
each relevant configuration of external quark flavours one or two of
the topologies shown in \reffi{fig:LOtops} contribute.
All LO and one-loop NLO diagrams are related by crossing symmetry to
the corresponding decay amplitude $\PH\to q\bar{q}q\bar{q}$. The QCD
and EW NLO corrections to these decays were discussed in detail in
\citeres{Bredenstein:2006rh,Bredenstein:2006ha}, in particular a
representative set of Feynman diagrams can be found there.

Evaluating $2\to3$ particle processes at the NLO level is
non-trivial, both in the analytical and numerical parts of the
calculation.  In order to ensure the correctness of our results we
have evaluated each ingredient twice, resulting in two completely
independent computer codes yielding results in mutual agreement.  The
phase-space integration is performed using multi-channel Monte Carlo
techniques \cite{Hilgart:1992xu} implemented in different ways in the
two different generators.

\subsection{Virtual corrections}

The virtual corrections modify the partonic processes that are
already present at LO; there are about 200 EW one-loop diagrams
per tree diagram in each flavour channel.
At NLO these corrections are induced by self-energy, vertex, 
box (4-point), and pentagon (5-point) diagrams.
The calculation of the one-loop diagrams has been performed in the
conventional 't~Hooft--Feynman gauge and in the background-field
formalism using the conventions of Refs.~\cite{Denner:1991kt} and
\cite{Denner:1994xt}, respectively. The masses of the external
fermions have been neglected whenever possible, i.e.\ everywhere but
in the mass-singular logarithms.

In the $s$-channel diagrams intermediate W and Z~bosons can become
resonant, corresponding to $\PW\PH/\PZ\PH$ production with subsequent
gauge-boson decay. In order to consistently include these resonances,
we implement the finite widths of the gauge bosons in the
``complex-mass scheme'', which was introduced in
Ref.~\cite{Denner:1999gp} for LO calculations and generalized to the
one-loop level in Ref.~\cite{Denner:2005fg}. In this approach the W-
and Z-boson masses are consistently considered as complex quantities,
defined as the locations of the propagator poles in the complex plane.
The scheme fully respects all relations that follow from gauge
invariance.

The amplitudes have been generated with {\sl FeynArts}, using the two
independent versions 1 \cite{Kublbeck:1990xc} and 3
\cite{Hahn:2000kx}.  The algebraic evaluation has been performed in
two completely independent ways. One calculation is based on the
in-house {\sl Mathematica} program that was already used in the
algebraic reduction of NLO corrections to the $\PH\to4\,$fermions
decays \cite{Bredenstein:2006rh,Bredenstein:2006ha}.  The other has
been completed with the help of {\sl FormCalc} \cite{Hahn:1998yk}.

The tensor integrals are evaluated as in the calculation of the
corrections to ${\rm e}^+{\rm e}^-\to4\,$fermions
\cite{Denner:2005fg,Denner:2005es}. They are recursively reduced to
master integrals at the numerical level. The scalar master integrals
are evaluated for complex masses using the methods and results of
Refs.~\cite{'tHooft:1978xw,Beenakker:1988jr,Denner:1991qq}. 
Tensor and scalar 5-point functions are directly expressed in terms of
4-point integrals \cite{Denner:2002ii}.  Tensor 4-point and 3-point
integrals are reduced to scalar integrals with the Passarino--Veltman
algorithm \cite{Passarino:1978jh} as long as no small Gram determinant
appears in the reduction. If small Gram determinants occur, the
alternative schemes described in Ref.~\cite{Denner:2005nn} are
applied. 

\subsection{Real corrections}

Real QCD corrections consist of gluon emission and processes
with $gq$ and $g\bar q$ initial states. Analogously real photonic
corrections comprise photon bremsstrahlung and photon-induced
processes with $\gamma q$ and $\gamma\bar q$ initial states.  The
matrix elements for these corrections have been evaluated using the
Weyl--van der Waerden spinor technique as formulated in
Ref.~\cite{Dittmaier:1998nn} and have been checked against results
obtained with {\sl Madgraph} \cite{Stelzer:1994ta}.

All types of real corrections involve singularities from collinear
initial-state splittings which are regularized with small quark
masses.  The mass singularities are absorbed via factorization by the
usual PDF redefinition both for the QCD and photonic corrections (see,
e.g., \citere{Diener:2005me}).  Technically, the soft and collinear
singularities for real gluon or photon emission are isolated both in
the dipole subtraction method following \citere{Dittmaier:1999mb} and
in the phase-space slicing method. For gluons or photons in the
initial state the subtraction and slicing variants described in
\citere{Diener:2005me} are applied. The results presented in the
following are obtained with the subtraction method, which numerically
performs better.


\section{Numerical results}

We use the input parameters as given in \citere{Bredenstein:2006rh}.
Since quark-mixing effects are suppressed, we set the CKM matrix to
the unit matrix.  The electromagnetic coupling is fixed in the $G_\mu$
scheme, i.e.\ it is set to $\alpha_{\GF}=\sqrt{2}\GF\/\MW^2\sw^2/\pi$,
because this accounts for electromagnetic running effects and some
universal corrections of the $\rho$ parameter.

We use the MRST2004QED PDF \cite{Martin:2004dh} which consistently
include ${\cal O}(\alpha)$ QED corrections. These PDF include a photon
distribution function for the proton and thus allow to take into
account photon-induced partonic processes.  As explained in
\citere{Diener:2005me}, the consistent use of these PDF requires the
$\MSbar$ factorization scheme for the QCD, but the DIS scheme for the
QED corrections; the corresponding factorization scales are identified
with the Higgs mass $\MH$ if not stated otherwise.  We only use four 
quark flavours for the
initial partons, i.e.\ we do not take into account the contribution of
bottom quarks, which is suppressed. Since no associated LO version of
the MRST2004QED PDF exists, we use these PDF both for LO and NLO
predictions.  For the renormalization scale of the strong coupling
constant by default we employ $\MH$, include 5 flavours in the
two-loop running, and fix $\als(\MZ)=0.1187$.

Apart from the total cross section without any phase-space cuts,
we consider the integrated cross section defined after applying
typical VBF cuts to the outgoing jets.
In this case, jets are defined from partons using the
$k_{\rT}$-algorithm \cite{Catani:1992zp,Catani:1993hr,Ellis:1993tq} 
as described in \citere{Blazey:2000qt}. 
More precisely, jets result from partons of
pseudorapidity $|\eta|<5$ using the jet resolution parameter 
$D=0.8$.  We also recombine real photons with partons or jets
according to this algorithm. Thus, some of the photons end up in jets,
others are left as identifiable photons.
Following \citere{Figy:2004pt}, we specifiy the VBF cuts as follows.
We require at least two hard jets with
\begin{equation}
\label{taggingjets}
p_{\rT\rj}\ge20\GeV,\qquad |y_{\rj}|\le4.5,
\end{equation}
where $p_{\rT\rj}$ is the transverse momentum of the jet and $y_{\rj}$
its rapidity. The tagging jets $\rj_1$ and $\rj_2$ are then defined as
the two jets passing the cuts \refeq{taggingjets} with highest
$p_{\rT}$ and $p_{\rT\rj_1}>p_{\rT\rj_2}$.  Finally, we demand a large
rapidity separation of the two tagging jets by
\begin{equation}
\Delta y_{\rj\rj}\equiv | y_{\rj_1}-y_{\rj_2}| > 4, \qquad y_{\rj_1}\cdot
y_{\rj_2}< 0.
\end{equation}

\begin{figure}
\includegraphics[bb= 60 445 280 635, scale=1.0]{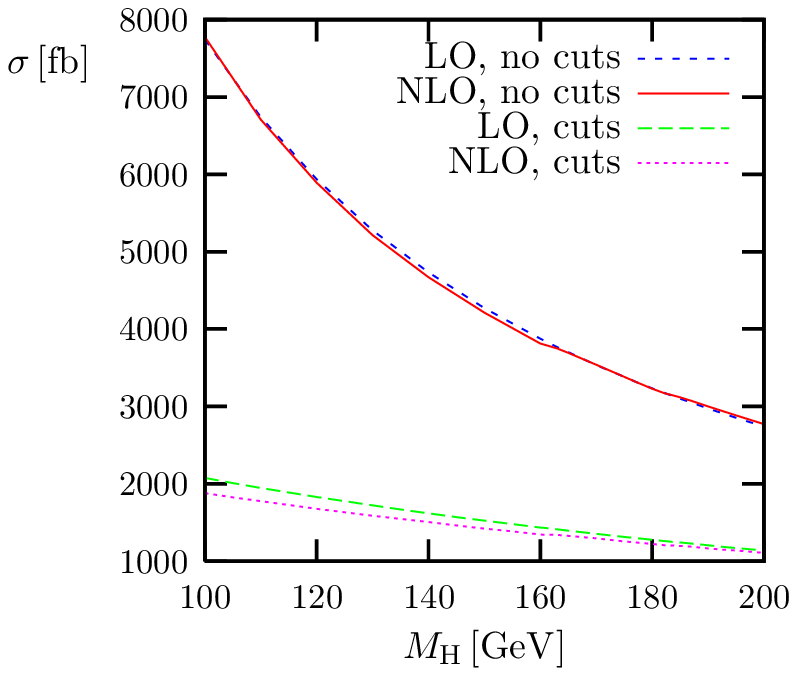}
\\[1ex]    
\includegraphics[bb= 60 445 280 635, scale=1.0]{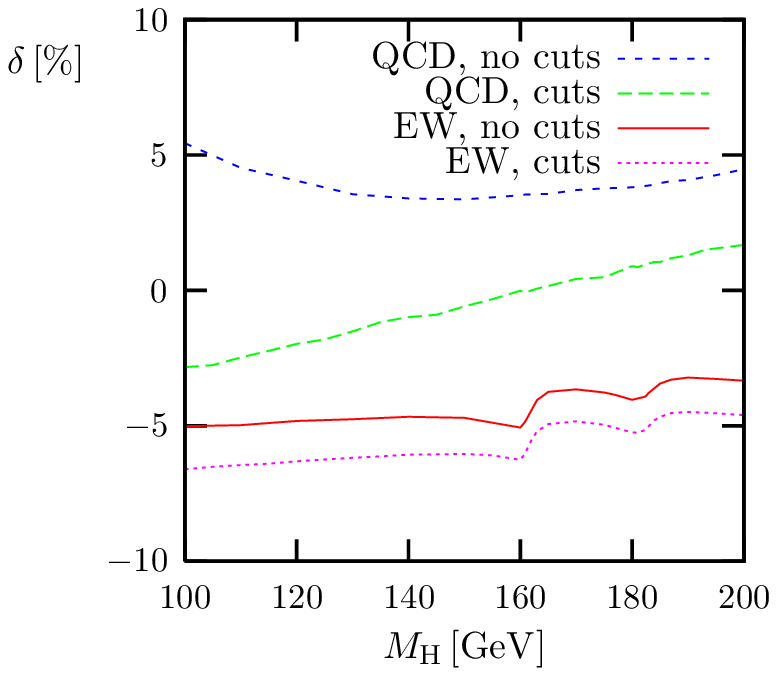}
\vspace*{-1em}
\caption{Higgs mass dependence of LO and complete NLO
  cross section (upper) and relative EW and QCD corrections (lower)
  without and with VBF cuts.}
\label{fi:mhdep}
\end{figure}

In \reffi{fi:mhdep} we plot the total cross section with and without
cuts as a function of $\MH$. In the upper panel we show the absolute
predictions in LO and in NLO including QCD and EW corrections. The VBF
cuts reduce the cross section by a factor 3--4. In the lower panel we
show the relative corrections. Without cuts the QCD corrections are
about +5\% and the EW corrections about $-5\%$ both depending only
weakly on $\MH$ and cancelling each other substantially.  With cuts
the EW corrections are approximately $-6\%$, while the QCD
corrections vary between $-3\%$ and $+2\%$. In the EW corrections the
WW and ZZ thresholds are clearly visible.
It is interesting to note that the EW corrections to the full VBF
channel are similar in size and sign to the EW corrections to the
subreactions $\Pp\Pp\to\PW\PH/\PZ\PH+X$ \cite{Ciccolini:2003jy}.
Compared to the related decays the $\PH\to\PW\PW/\PZ\PZ\to4f$
\cite{Bredenstein:2006rh,Bredenstein:2006ha} the size is similar, but
the sign is different.

In \refta{ta:xsection_nocuts} we present integrated cross
sections for $\MH=120$, 150, 170, and $200\GeV$ without any cuts and
in \refta{ta:xsection_cuts} results for VBF cuts.  We list the LO
cross section $\sigma_{\mathrm{LO}}$, the cross section
$\sigma_{\mathrm{NLO}}$ including QCD+EW corrections, and the
relative QCD and EW corrections, $\delta_{\mathrm{QCD}}$ and
$\delta_{\mathrm{EW}}$, respectively.  The complete EW corrections
$\delta_{\mathrm{EW}}$ also comprise the corrections from
photon-induced processes $\delta_{\gammainduced}$, which turn out to
be $\sim+1\%$ and to reduce the EW corrections.
We note that the QCD corrections are dominated
by the known (vertex-like) corrections to the squared $t$- and
$u$-channel VBF diagrams, while corrections to interference terms are
at the level of
$0.1\%$.%
\begin{table}
\def\phm{\phantom{-}}
\def\phn{\phantom{0}}
\centerline{
\begin{tabular}{|c|c|c|c|c|}
\hline
$\MH\ [\GeV]$ & 120 & 150 & 170 & 200 \\
\hline
$\sigma_{\mathrm{LO}}\ [\fb]$
& 5936(1)  
& 4271(2)  
& 3536(1)  
& 2743(1)  
\\
$\sigma_{\mathrm{NLO}}\ [\fb]$
& 5890(2)  
& 4219(2)  
& 3538(1)  
& 2775(1)  
\\
$\delta_{\mathrm{QCD}}\ [\%]$
& $\phm4.04(3)$  
& $\phm3.47(2)$  
& $\phm3.72(2)$  
& $\phm4.48(2)$  
\\
$\delta_{\mathrm{EW}}\ [\%]$
& $-4.81(2)$  
& $-4.70(2)$  
& $-3.65(1)$  
& $-3.33(1)$  
\\
$\delta_{\gammainduced}\ [\%]$
& $\phm0.86(1)$  
& $\phm1.04(1)$  
& $\phm1.14(1)$  
& $\phm1.27(1)$  
\\
\hline
\end{tabular}
}
\caption{Cross section for $\ppjjh$ in LO and NLO without cuts
  and relative QCD and EW corrections. The contribution
  $\delta_{\gammainduced}$ from $\gamma$-induced processes (which
  is part of $\delta_{\mathrm{EW}}$) is also given separately.}
\label{ta:xsection_nocuts}
\end{table}
\begin{table}
\def\phm{\phantom{-}}
\def\phn{\phantom{0}}
\centerline{
\begin{tabular}{|c|c|c|c|c|}
\hline
$\MH\ [\GeV]$ & 120 & 150 & 170 & 200 \\
\hline
$\sigma_{\mathrm{LO}}\ [\fb]$
& 1830.5(5) 
& 1524.2(4) 
& 1353.8(3) 
& 1139.1(3) 
\\
$\sigma_{\mathrm{NLO}}\ [\fb]$
& 1678.7(9) 
& 1422.9(7) 
& 1293.4(6) 
& 1106.0(5) 
\\
$\delta_{\mathrm{QCD}}\ [\%]$
& $-1.97(4)$  
& $-0.60(4)$  
& $\phm0.41(4)$ 
& $\phm1.76(3)$ 
\\
$\delta_{\mathrm{EW}}\ [\%]$
& $-6.32(2)$  
& $-6.02(2)$  
& $-4.87(1)$  
& $-4.64(1)$  
\\
$\delta_{\gammainduced}\ [\%]$
& $\phm1.14(1)$  
& $\phm1.21(1)$  
& $\phm1.25(1)$  
& $\phm1.31(1)$  
\\
\hline
\end{tabular}
}
\caption{As in \refta{ta:xsection_nocuts}, but with VBF cuts applied.}
\label{ta:xsection_cuts}
\end{table}
These interference corrections are not enhanced by contributions of two
$t$- or $u$-channel vector bosons with small virtuality and therefore even
further suppressed when applying VBF cuts.  

\begin{figure}
\includegraphics[bb= 60 445 280 635, scale=1]{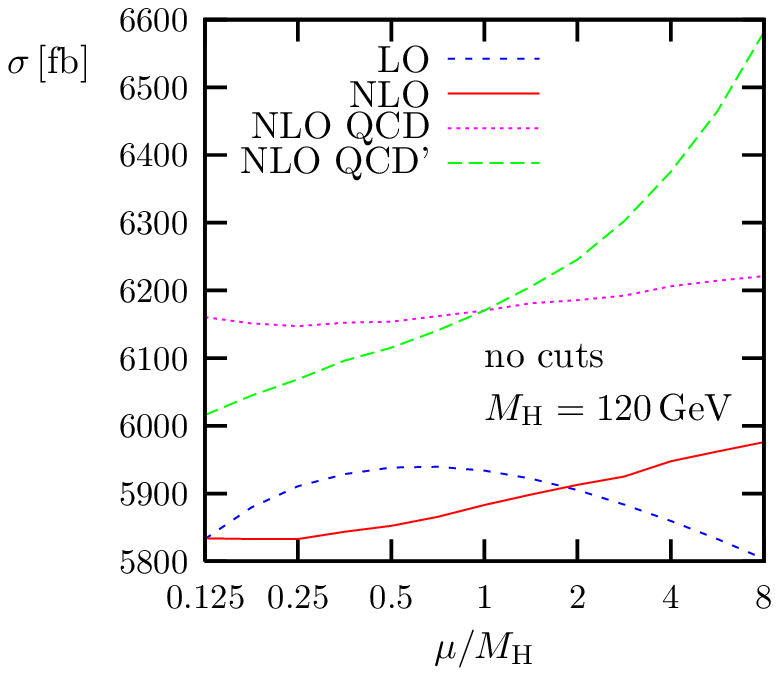}
\\[2ex]    
\includegraphics[bb= 60 445 280 635, scale=1]{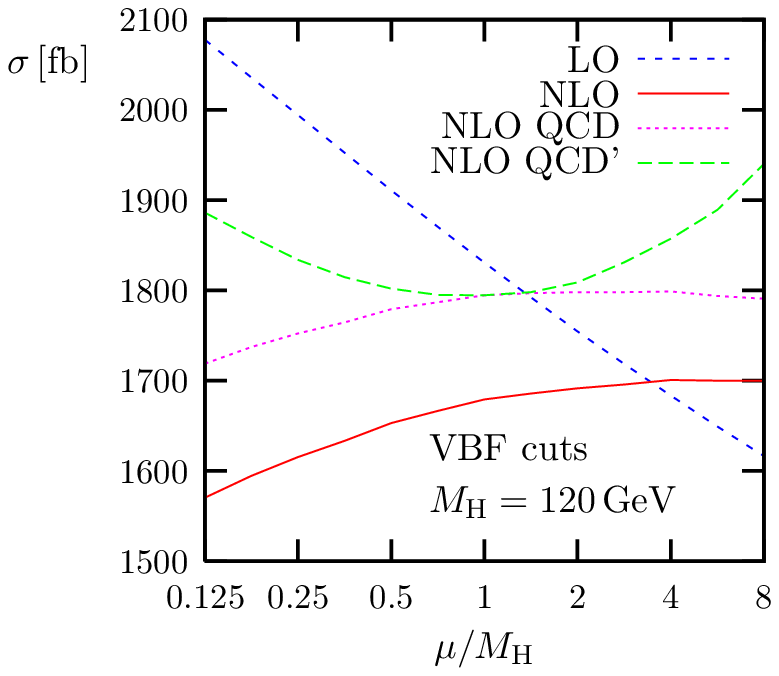}
\vspace*{-1em}
\caption{Scale dependence of LO and NLO cross sections with QCD or QCD+EW
  corrections for $\MH=120\GeV$ without cuts (upper)
  and with VBF cuts (lower), $\mu_{\mathrm{R}}=\mu_{\mathrm{F}}\equiv\mu$
  for LO, NLO and NLO QCD, but $\mu_{\mathrm{R}}=\MH^2/\mu$ for NLO QCD'.}
\label{fi:mudep_120}
\end{figure}

In \reffi{fi:mudep_120} we show the dependence of the total cross
section on the factorization and renormalization scale for
$\MH=120\GeV$.  We set the factorization scale
$\mu\equiv\mu_{\mathrm{F}}$, which applies to both QCD and QED
contributions, equal to the renormalization scale
$\mu_{\mathrm{R}}=\mu$ and vary it between $\MH/8$ and $8\MH$.  In
this set-up, we show the LO cross section, the QCD corrected NLO cross
section and the complete NLO cross section involving both QCD and EW
corrections.  In addition we depict the QCD corrected NLO cross
section for the setup where $\mu_{\mathrm{R}}=\MH^2/\mu$ (NLO QCD').
Varying the scale $\mu$ up and down by a factor $2$ ($8$) changes
the cross section by 11\% (29\%) in LO and 3\% (18\%) in NLO for the
set-up with VBF cuts.  Without cuts the scale uncertainty it is at the
level of 3\% (11\%) in NLO, while it is accidentally small in LO for
this specific Higgs-boson mass.

\section{Conclusions}

Radiative corrections of strong and electroweak interactions have been
discussed at next-to-leading order for Higgs production via
vector-boson fusion at the LHC.  The electroweak corrections, which
have not been calculated before, reduce the cross section by $5\%$,
and are thus as important as the QCD corrections in this channel.  QCD
corrections to interference contributions turn out to be negligible,
confirming previous approximations.
\\[1em]
Acknowledgement: This work is supported in part by the European
Community's Marie-Curie Research Training Network HEPTOOLS under
contract MRTN-CT-2006-035505. 
We thank M. Spira for comments on the manuscript.

\bibliography{ppjjh_let}

\end{document}